\documentclass[aps,prl,amsmath,amssymb,superscriptaddress,twocolumn,showpacs,longbibliography,floatfix,notitlepage]{revtex4-1}

\DeclareSymbolFontAlphabet{\amsmathbb}{AMSb}%
\DeclareSymbolFontAlphabet{\mathbb}{AMSb}
\usepackage[bbgreekl]{mathbbol}

\usepackage{graphicx}
\usepackage{dcolumn}
\usepackage[colorlinks=true,linkcolor=blue ,citecolor=blue,urlcolor=blue]{hyperref}
\usepackage{multirow}
\usepackage[usenames,dvipsnames]{xcolor}
\usepackage{soul}
\usepackage{braket}
\usepackage{bm}
\usepackage{nicefrac}
\usepackage{siunitx}
\usepackage{color}
\usepackage[utf8]{inputenc}   
\usepackage{upgreek}
\usepackage{ulem}
\usepackage{float}

\newcommand{\VEC}[1]{\boldsymbol{#1}}

\begin{document}

\title{Fermi-surface origin of  skyrmion lattices in centrosymmetric rare-earth intermetallics}

\author{Juba Bouaziz}
\email{j.bouaziz@fz-juelich.de}
\affiliation{Peter Gr\"{u}nberg Institut and Institute for Advanced Simulation, Forschungszentrum J\"{u}lich \& JARA, D-52425 J\"{u}lich, Germany}
\affiliation{Department of Physics, University of Warwick, Coventry CV4 7AL, United Kingdom}
\author{Eduardo Mendive-Tapia}
\affiliation{Peter Gr\"{u}nberg Institut and Institute for Advanced Simulation, Forschungszentrum J\"{u}lich \& JARA, D-52425 J\"{u}lich, Germany}
\affiliation{Department of Computational Materials Design, Max-Planck-Institut für Eisenforschung, 40237 Düsseldorf, Germany}
\author{Stefan Bl\"ugel}
\affiliation{Peter Gr\"{u}nberg Institut and Institute for Advanced Simulation, Forschungszentrum J\"{u}lich \& JARA, D-52425 J\"{u}lich, Germany}
\author{Julie B. Staunton}
\affiliation{Department of Physics, University of Warwick, Coventry CV4 7AL, United Kingdom}

\date{\today}

\begin{abstract}
{We show from first-principles that  barrel-shaped  structures within the  Fermi surface of the centrosymmetric intermetallic compounds GdRu$_2$Si$_2$ and Gd$_2$PdSi$_3$ give rise to Fermi surface nesting, which determines the strength and sign of quasi-two-dimensional Ruderman-Kittel-Kasuya-Yosida pairwise exchange  interactions between the Gd moments. This is  the principal mechanism leading to their helical single-$q$ spin-spiral ground states, providing transition temperatures and magnetic periods in good agreement with experiment. Using atomistic spin-dynamic simulations, we draw a direct line between the subtleties of the three-dimensional Fermi surface topology and the stabilization of a square skyrmion lattice in GdRu$_2$Si$_2$  at applied magnetic fields  as observed in experiment.}

\end{abstract}
 
\maketitle
Magnetic skyrmions have been under intense scrutiny over the last decade, mainly owing to their intriguing topological and transport properties~\cite{Neubauer:2009}, and prospect as entities for  ultra-low power information storage and processing devices~\cite{nagaosa2013topological,fert2013skyrmions}.
Hexagonal skyrmion crystals (SkX) and single skyrmions are prevalent in non-centrosymmetric materials where they emerge from the competition between ferromagnetic exchange and the Dzyaloshinskii-Moriya interaction (DMI) in the presence of small applied magnetic fields~\cite{roessler2006spontaneous,nagaosa2013topological}. The small energy scale of the relativistic DMI gives rise to skyrmions with typical sizes of tenths of nanometers or more.
A path to still smaller skyrmions is provided by the replacement of the DMI by much larger antiferromagnetic exchange interactions competing with ferromagnetic ones on equal footing. A control of skyrmion formation by small  applied fields is, therefore, not trivial~\cite{Nandy2016}.  

More flexibility, which does not rely on DMI, can be obtained by exploiting intrinsically competing exchange interactions, such as Ruderman-Kittel-Kasuya-Yosida (RKKY) interactions, prevalent in centrosymmetric rare-earth intermetallics~\cite{hughes2007lanthanide,Mendive:2017}. Here the interaction between a pair of \textit{e.g.}\ Gd atoms oscillates between being ferro- or antiferromagnetic depending on the magnitude of the position vector, ${\bf R}_{1}-{\bf R}_{2}$, connecting the two atoms and its  orientation within  the lattice.  The mechanism and strength  is fundamentally linked to the topology and nesting features of the Fermi surface (FS)~\cite{mackintosh1991rare,hughes2007lanthanide}, the symmetry of the crystal, and the conduction electrons that own the FS. Based on a model for conduction electrons described by a nearly isotropic two-dimensional (2D) electron gas, Wang \textit{et al}.~\cite{Wang:2020} indeed pointed out the possibility  of  stabilizing a hexagonal skyrmion crystal (SkX) phase by the RKKY interaction. 

Recently, skyrmion crystals  with small lattice constants  were experimentally uncovered for two different centrosymmetric ternary Gd-$4d$-transition-metal intermetallic compounds:  Gd$_2$PdSi$_3$, formed by 
stacked hexagonal Gd layers, exhibiting a hexagonal SkX with a  lattice constant of  $\sim 2.5$~nm~\cite{kurumaji2019skyrmion}, and GdRu$_2$Si$_2$, where 2D sheets of  Gd atoms arranged on a square lattice separated by layers of Ru and Si atoms exhibit a square lattice of skyrmions, a so-called double-$q$ SkX, with an even smaller lattice constant of $1.9$~nm ~\cite{khanh2020nanometric}. The SkX phases with their small lattice constants display many interesting properties, \textit{e.g.}\ a giant topological Hall effect~\cite{kurumaji2019skyrmion,Bouaziz2021}, a large topological Nernst~\cite{Hirschberger:2020}, and a planar Hall effect~\cite{Hirschberger_2:2020}. 

In contrast  to the SkX formation suggested by the RKKY model of Wang \textit{et al}.\,\cite{Wang:2020}, a recent theoretical in-depth analysis fundamentally  rejected the  RKKY  mechanism for both compounds and proposed an inter-orbital Gd $d$-$f$ frustration as direct exchange mechanism between Gd atoms~\cite{Nomoto:2020}.

In this letter, we  demonstrate fully from first-principles, that the RKKY paradigm is the central mechanism in  GdRu$_2$Si$_2$ and Gd$_2$PdSi$_3$ giving rise to a nanometric single-$q$ spiral state. This spiral forms within the Gd planes, weakly ferromagnetically coupled along the $c$-direction, and transforms to a SkX under an applied magnetic field.  We show how the formation of a three-dimensional (3D) barrel-shaped FS topology around the $\Gamma$ point and its subsequent nesting directly coincides with the {large formation energy and the wave vector $\VEC{q}$ of the spin spiral.} 
The 3D itinerant valence electron glue  of primarily $4d$ Ru or Pd electrons hybridizing with the $5d$ electrons of Gd {mediates the magnetic interactions between the Gd atoms}. We calculate the transition temperatures, finding a good quantitative agreement with experiments~\cite{kurumaji2019skyrmion,khanh2020nanometric}.

%%%%%%%%%%%%%%%%%%%%%%%%%%%%%%%%%%%%%%%%%%%%%%%%%%%%%%%%%%%%%%%%%%%%
\begin{figure*}
  \centering
  \includegraphics[width=1.0\textwidth]{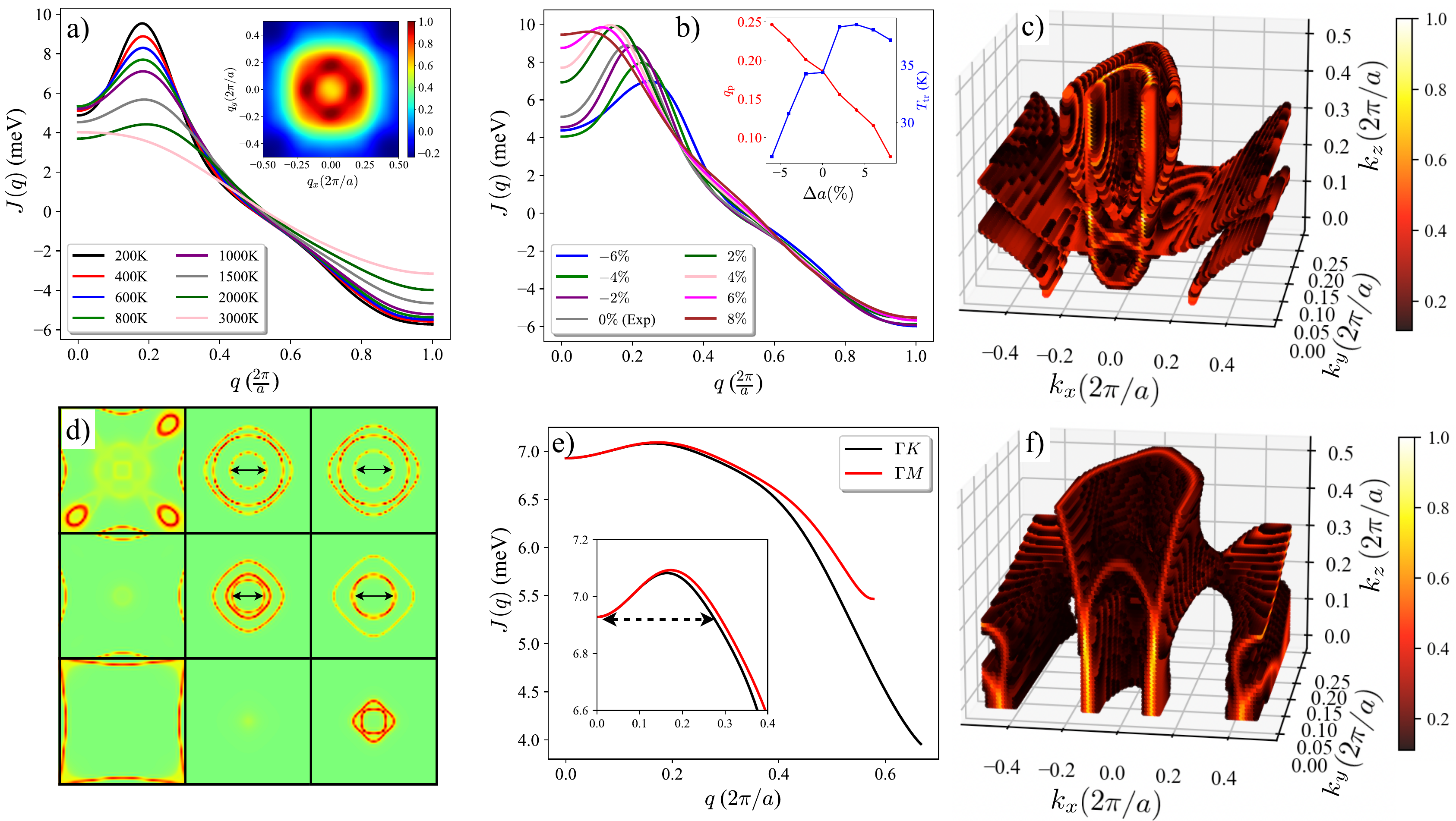}
  \caption{
  	a) Fourier transform, $J(\boldsymbol{q})$, of the magnetic interactions for different electronic  temperatures at the experimental lattice parameter $a_\text{exp}=4.162$~\AA\ of GdRu$_2$Si$_2$ against the wave vector $\boldsymbol{q}$ along the [100] direction. The inset shows a colormap encoding the size of $J(\boldsymbol{q})$ within the $q_z=0$ plane. b) $J(\boldsymbol{q})$ of GdRu$_2$Si$_2$ for different lattice compressions and expansions.
  	c,f) Fermi surfaces of GdRu$_2$Si$_2$ and Gd$_2$PdSi$_3$, respectively, where the color indicates the intensity of the spectral weight $A_\mathrm{B}(\VEC{q},E)/\max_{\VEC{q},E}A_\mathrm{B}(\VEC{q},E)$. 
  	d) Cuts of Fermi surface of GdRu$_2$Si$_2$ at $k_z\in\{0, 0.3, 0.5\}\frac{2\pi}{a}$, from left to right, with $a=a_\text{exp}-6\%$ (upper panels),  $a=a_\text{exp}$ (middle panels), and  $a=a_\text{exp}+8\%$  (lower panels). e) $J(\boldsymbol{q})$ of Gd$_2$PdSi$_3$ along the $\Gamma K$ and $\Gamma M$ directions.
  	}
  \label{panel_1}
\end{figure*}
%%%%%%%%%%%%%%%%%%%%%%%%%%%%%%%%%%%%%%%%%%%%%%%%%%%%%%%%%%%%%%%%%%%%

We go on to investigate the effect of an external magnetic field, $B$, applied to GdRu$_2$Si$_2$. We show that the lowest energy state at intermediate magnetic fields corresponds to a 2$q$, square  SkX that arises {directly} from the combination of crystal symmetry-dependent frustrated RKKY interactions with a {simple uniaxial} magnetocrystalline anisotropy and a magnetic field perpendicular to the helical axis. While four-spin interactions can favor the stability of multiple-$q$ states and square  SkX, as explained in Refs.~\cite{Kurz2001,khanh2020nanometric,Heinze2011}, our results demonstrate that multisite interactions are not necessary to produce the  square SkX in GdRu$_2$Si$_2$. We also show that the SkX phase exhibits a strong  magneto-elastic coupling.

%The first-principles calculations are carried out employing the Green function-based formulation of density functional theory using the Korringa-Kohn-Rostoker method~\cite{papanikolaou:2002,Bauer:2014}. The generalized gradient approximation (GGA) is used for the exchange-correlation energy~\cite{Perdew:1996}. We use the infinitesimal rotation method for the computation of pairwise magnetic interactions~\cite{liechtenstein1984,Ebert:2009,Solovyev:2021} between localized Gd moments, $\{J_{ij}\}$, evaluated in the ferromagnetic state. See the Supplemental Material for details~\footnote{See Supplemental Material, which contains references \cite{PhysRevB.99.144424,elliott2013magnetic,PhysRevB.5.2382,PhysRev.156.809,Hayami2021}, for further details on the Fourier transform of the magnetic interactions and the construction of the magnetic phase diagram. We also show additional plots for Fermi surface cuts of Gd$_2$PdSi$_3$, the density of states of the two materials studied, components of spin structure factors, and a list of material families which may host skyrmions.}.

%The first-principles calculations are carried out employing
We employ the Green function-based formulation of density functional theory using the scalar relativistic, all-electron full-potential Korringa-Kohn-Rostoker (KKR) method~\cite{papanikolaou:2002,Bauer:2014}. The exchange-correlation energy is treated in the generalized gradient approximation (GGA)~\cite{Perdew:1996}. The magnetic pair interactions, $\{J_{ij}\}$, are computed using the infinitesimal rotation method in the ferromagnetic state~\cite{liechtenstein1984,Ebert:2009,Solovyev:2021}.

\textit{RKKY mechanism and Fermi surface nesting in  GdRu$_2$Si$_2$ and Gd$_2$PdSi$_3$} --
Taking the lattice Fourier transform of $\{J_{ij}\}$ gives the Gd-Gd interaction in wave vector (momentum) space, $J(\boldsymbol{q})$~\cite{Mendive:2021} (see Supplemental Material for details~\footnote{See Supplemental Material, which contains references \cite{PhysRevB.99.144424,elliott2013magnetic,PhysRevB.5.2382,PhysRev.156.809,Hayami2021}, for further details on the Fourier transform of the magnetic interactions and the construction of the magnetic phase diagram. We also show additional plots for Fermi surface cuts of Gd$_2$PdSi$_3$, the density of states of the two materials studied, components of spin structure factors, and a list of material families which may host skyrmions.}). The highest value of $J(\boldsymbol{q})$ determines {the N\'eel temperature and} the wave vector $\boldsymbol{q}_p$ of the spiroidal magnetic single-$q$ state of lowest energy, while the energy difference, $J(\boldsymbol{q}_p) - J(\boldsymbol{0})$, describes its stability with respect to the ferromagnetic one.
Fig.\ \ref{panel_1}(a) shows $J(\boldsymbol{q})$ along the [100] direction for GdRu$_2$Si$_2$ evaluated using the experimental lattice parameter. It features a typical paramagnetic spin susceptibility of a metal with a FM ground state augmented by a sharp, significant finite-q peak at $q_p=0.19\,\frac{2\pi}{a}$. This is consistent with the low temperature stabilization of an incommensurate helimagnetic state,
in very good agreement with the experiment,
$q_p^\text{exp}=0.22\, \frac{2\pi}{a}$~\cite{khanh2020nanometric}. At low levels of thermally-induced single electron-hole excitations across the FS (e.g. electronic temperature of 200\,K), $J(\boldsymbol{q}_p)=9.5$\,meV, nearly twice the size of $J(\boldsymbol{0})=4.9$\,meV.
This peak is consistent to the long-range magnetic interactions that extend somewhat further than three lattice constants, in accordance with an RKKY picture~\cite{Bouaziz2017}. For example, if we only consider shorter interactions (below $3a$), we observe that $\boldsymbol{q}_p = 0$, and so a ferromagnetic state stabilizes instead.
%The momentum-dependent Gd-Gd interaction,
$J(\boldsymbol{q})$ can be conveniently analysed in terms of a Fermi sea contribution to the 3D bulk magnetic susceptibility, $J_\textrm{3D}(\boldsymbol{q})$, and the FS specific contribution to the RKKY interaction, $J_\mathrm{FS}(\boldsymbol{q})$, $J(\boldsymbol{q})=J_\mathrm{3D}(\boldsymbol{q})+J_\mathrm{FS}(\boldsymbol{q})$. To distinguish $J_\mathrm{3D}$ from $J_\mathrm{FS}$ and provide  further evidence for this RKKY mechanism with its fundamental link to the FS,  the effect of the electronic temperature $T$ prescribed by the Fermi-Dirac distribution for the single electron-hole excitations is investigated. In Fig.~\ref{panel_1}(a), we demonstrate how the peak vanishes when $T$ increases, which shows how the stabilization of the spiral state is strongly tied to the itinerant valence electronic structure.
We also highlight that $J(\boldsymbol{q})$ peaks along all directions within the $\boldsymbol{q}=(q_x,q_y,0)$ plane with roughly the same size of the wave vector $q_p$. The peak height is modulated  and thus the strength of the RKKY interaction varies  in accordance with the $C_{4v}$ symmetry of  the crystal lattice. This is shown in the inset of Fig.\ \ref{panel_1}(a), which is a color map for the value of $J(\boldsymbol{q})$ in the $q_z=0$ plane.
No peak is observed in the out-of-plane directions.

Close scrutiny of the FS reveals how  its topology is connected to the emergence of $\boldsymbol{q}_p$ {and the size of $J(\boldsymbol{q}_p)$}.
Fig.~\ref{panel_1}(c,f) shows the 3D FS of GdRu$_2$Si$_2$ as well as of Gd$_2$PdSi$_3$, where the color indicates the spectral weight~\cite{ebert2011calculating}. These calculations are suitably performed in the paramagnetic state, which we describe by fully disordered local moments (DLM) averaged using the coherent potential approximation~\cite{pindor1983,staunton1984,gyorffy1985}. We have found in both materials the presence of barrel-shaped nested sheets with a large spectral weight located around the $\Gamma$ point. Their correlation with  $\boldsymbol{q}_p$ in GdRu$_2$Si$_2$ can be seen by examining FS  cuts at $k_z\in\{0, 0.3, 0.5\}\frac{2\pi}{a}$, see Fig.~\ref{panel_1}(d). For the experimental lattice parameter (middle panels), we observe that the magnitude of the wave vector of the magnetic spiral, $\boldsymbol{q}_p$, matches the nesting vector $\VEC{k}_\mathrm{N}$, \textit{i.e.}\ the Bloch wave vector spanning the  barrel-like nesting in the ($k_x$,$k_y$) plane {and showing $C_{4v}$ symmetry of the crystal}. The length of {$\VEC{k}_\mathrm{N}$}, indicated by a black arrow {($\leftrightarrow$)}, increases  (reduces/disappears) when the in-plane lattice parameter $a$ decreases (increases), keeping the out-of-plane lattice constant fixed, as shown in the upper (lower) panels. We observe that this directly correlates with the size of $\boldsymbol{q}_p$ plotted in Fig.~\ref{panel_1}(b) for different values of $a$. In other words, the {barrel} collapses when the peak vanishes  [see Fig.~\ref{panel_1}(d)].

The Fourier transform of the magnetic interactions described by an RKKY mechanism is directly proportional to the paramagnetic susceptibility of the valence electrons, dominated by the convolution of the Bloch spectral functions, $A_\mathrm{B}(\boldsymbol{k},E)$, at the Fermi energy, $E_\mathrm{F}$, at low temperature
\begin{equation}
J_\mathrm{FS}(\boldsymbol{q}) \approx  \int_\mathrm{BZ} d\boldsymbol{k}\,A_\mathrm{B}(\boldsymbol{k},E_\text{F})A_\mathrm{B}(\boldsymbol{k}+\boldsymbol{q},E_\text{F})
    \label{EQNest}\, .
\end{equation}
$J_\mathrm{FS}(\boldsymbol{q})$, therefore, should
present a peak at $\VEC{q}_p$ when the FS contains nesting features~\cite{mackintosh1991rare}. 
{Making an ansatz of a 2D electron dispersion $E(k_x,k_y)$, Wang \textit{et al}.\,\cite{Wang:2020} showed the presence of a small nesting induced  peak of  $J(\boldsymbol{q})$. However, if the nesting  condition holds  during the integration along the third dimension $k_z$ in the 3D  barrel-shaped FS as shown in Fig.~\ref{panel_1}(c,d,f), the peak of  $J_\mathrm{FS}(\boldsymbol{q}_p)\approx J(\boldsymbol{q}_p)-J(\boldsymbol{0})\propto H$  scales with the height $H$ of the barrel}.  

Also for Gd$_2$PdSi$_3$ a similar pipe-like shape in the FS  forms with a nesting vector that is very close to $q_p=0.17\,\frac{2\pi}{a}$ (in the $\Gamma M$ direction)  as shown in Fig.~\ref{panel_1}(e,f). This value is also in good agreement with experiment, $q_p^\text{exp}=0.14\,\frac{2\pi}{a}$~\cite{kurumaji2019skyrmion, Inosov2009}. The set of $\{J_{ij}\}$, and $J({\boldsymbol{q})}$,  inherit the hexagonal symmetry of this compound such that the $\boldsymbol{q}$-vector of the spiral wave follows the high-symmetry direction of the Gd lattice.
The strongest hexagonal warping is present for a nesting vector $k_\mathrm{N}\approx q_p$, from which the finite peak originates in $J(\VEC{q})$. The size of the nesting vector, {${k}_\mathrm{N}(k_z)$} depends on {$k_z$}, as shown in Fig.~\ref{panel_1}(f). The largest length is  $k_\mathrm{N}\approx 0.4\frac{2\pi}{a}$ directly relates to the width of the peak of $J(\boldsymbol{q})$ as indicated by a black dashed line in Fig.~\ref{panel_1}(e). 

\textit{Transition temperature} -- A mean-field estimation of the Néel transition temperature from the paramagnetic state is provided by $T_\text{N}=\frac{J(\boldsymbol{q}_p)}{3k_\text{B}}$, $k_\text{B}$ being the Boltzmann constant. For the corresponding experimental lattice parameters we have obtained $T_\text{N}=34$\,K, a spiral period of ($\lambda={2\pi}/{q_p}$) $\lambda=2.19$\,nm and a magnetic moment of $M_\mathrm{Gd}=7.048$\,$\mu_\mathrm{B}$ for GdRu$_2$Si$_2$, and $T_\text{N}= 27$\,K, {$M_\mathrm{Gd}=7.057$\,$\mu_\mathrm{B}$,} and $\lambda=4.77$\,nm for Gd$_2$PdSi$_3$.
The computed transition temperatures are in excellent agreement with experiment, $T_\text{N}^\text{exp}=46$\,K (GdRu$_2$Si$_2$) and $T_\text{N}^\text{exp}=20$\,K (Gd$_2$PdSi$_3$), and the theoretical spiral periods correctly lie within the experimental nanometer scale, $\lambda^\text{exp}=1.90$\,nm (GdRu$_2$Si$_2$)~\cite{khanh2020nanometric} and $\lambda^\text{exp}=5.82$\,nm (Gd$_2$PdSi$_3$)~\cite{kurumaji2019skyrmion}. 

\textit{Magneto-elastic coupling --} In the inset of Fig.~\ref{panel_1}(b), we show $q_p$ and the transition temperature of GdRu$_2$Si$_2$ as functions of the in-plane lattice parameter $a$ while keeping $c$ fixed. A lattice compression of $-6.0\%$ enhances $q_p$ from $0.19$\,$\frac{2\pi}{a}$ to $0.25$\,$\frac{2\pi}{a}$ whereas an expansion progressively reduces {the energy difference $J(\boldsymbol{q}_p) - J(\boldsymbol{0})$ and the wave vector}  $q_p$ to reach almost zero at  $8\%$. In contrast the Néel temperature, $T_\text{N}$, increases (decreases) for lattice expansion (contraction) and ranges
over $24-38$\,K.

%%%%%%%%%%%%%%%%%%%%%%%%%%%%%%%%%%%%%%%%%%%%%%%%%%%%%%%%%%%%%%%%%%%%
\begin{figure}
 \centering
  \includegraphics[width=0.49\textwidth]{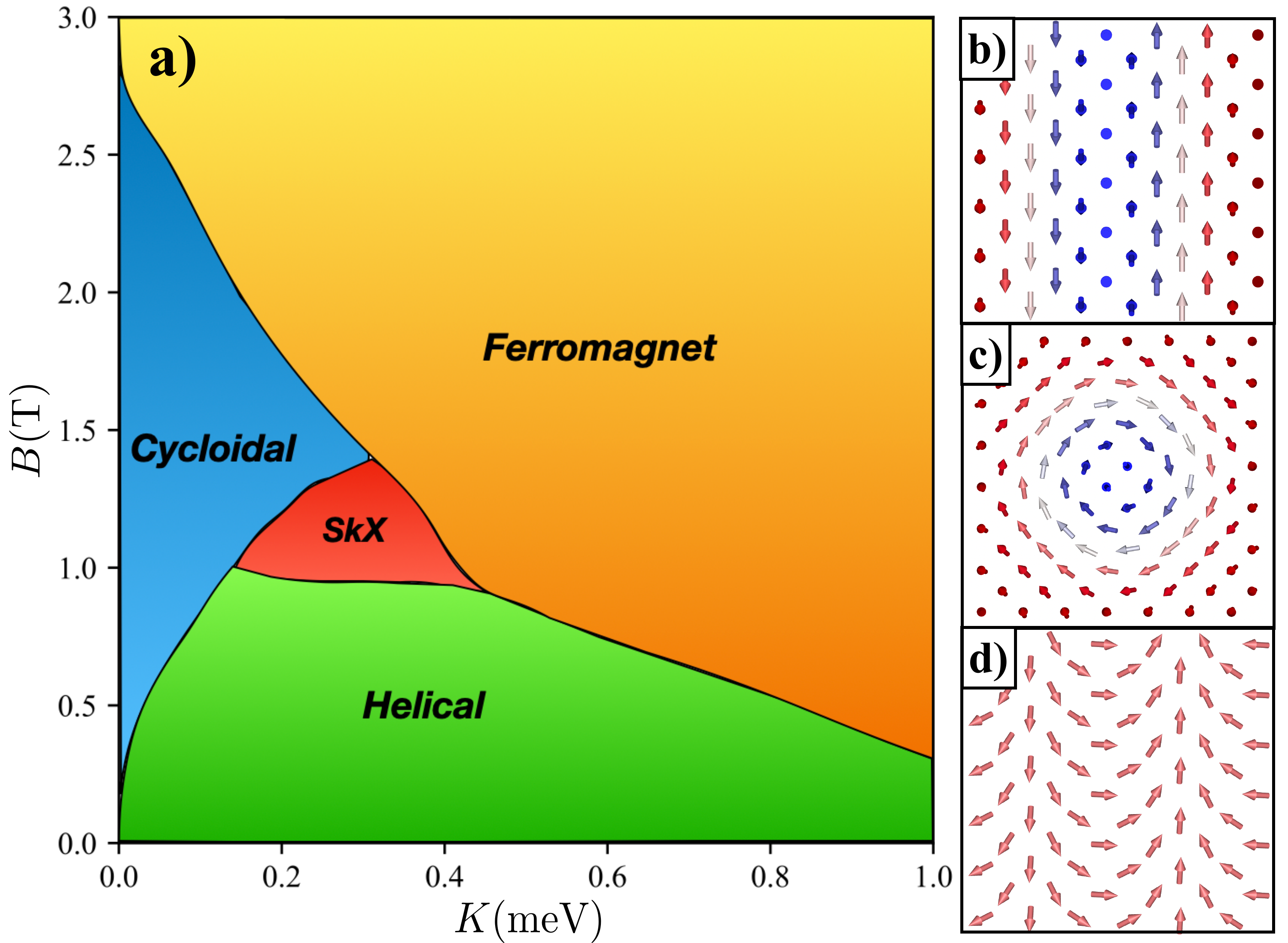}
  \caption{
  a) GdRu$_2$Si$_2$ magnetic phase diagram exhibiting four different phases in the ($K, B$) parameter space. $K$ is magneto-crystalline anisotropy and $B$ is the magnetic field. (b-d) Representative magnetic textures (color indicates the $m_z$ component, blue: $m_z=-1$, red $m_z=1$) described by the Gd moments at lattice sites bct crystal structure corresponding to: (b) helical state  for $(K,B)=(0.2,0.6)$, (c) SkX for $(K,B)=(0.2,1.0)$ and (d) cycloidal for $(K,B)=(0.1, 1.0)$.}
  \label{panel_ph}
\end{figure}
%%%%%%%%%%%%%%%%%%%%%%%%%%%%%%%%%%%%%%%%%%%%%%%%%%%%%%%%%%%%%%%%%%%%

\textit{Skyrmion crystal} --- {We now turn our attention to} the stabilization of a SkX phase {in GdRu$_2$Si$_2$} upon 
application of a magnetic field. Experimentally, it has been reported that the wave vector $\VEC{q}_p$ is normal to the plane of the helimagnetic 
state at zero magnetic field~\cite{devishvili2010magnetic,khanh2020nanometric}. This magnetic 
state is in line with an easy $c$-axis magnetic anisotropy as reported in Ref.~\cite{garnier1995anisotropic}. 
The formation of a 2$q$-SkX  state is found when an external magnetic field is applied 
along the $c$-axis, \textit{i.e.} parallel to the magnetization plane of the helimagnetic state. As the applied $B$-field is increased, the helix is distorted until the field reaches $\simeq2$\,T, at which a phase transition to the SkX state occurs.
This SkX state remains stable for temperatures up to $\simeq20$\,K. 
When the field is increased further, initially a cycloid magnetic phase emerges, which then changes to a saturated ferromagnetic state~\cite{khanh2020nanometric}.

A natural path to identify the magnetic phases present in GdRu$_2$Si$_2$ is to analyse a micromagnetic energy functional. By fitting  $J({q})$ for small $q$, as $J({q})\simeq \mathcal{A}q^2 +\mathcal{B}q^4 + \mathcal{C}q^2_xq^2_y$ ($\mathcal{A}$ to $\mathcal{C}$ being constants) and then taking the continuum limit, we identify higher-power gradient terms such as  $q^4\leftrightarrow (\nabla^2\VEC{m})^2$ of the magnetization density $m$ as the stabilization mechanism of a SkX~\cite{rybakov2019magnetic,grytsiuk2020}. However, to cover the short magnetic wavelengths, even higher powers of $q$ would be required in the fit. {Therefore, to tackle this complexity more effectively, we adopt} 
an atomistic description where we use directly the magnetic interactions which we have computed from first-principles (see Fig.~\ref{panel_1}). We use the internal energy: 

\begin{equation}
\begin{split}
{\mathcal{E}_{m} =  -\frac{1}{2}\sum_{ij}J_{ij}\VEC{m}_{i}\VEC{m}_{j}
-K\sum_{i}(\VEC{m}_{i}\cdot\VEC{e}_z)^2
-\VEC{B}\cdot\sum_i\VEC{m}_{i},
}
\end{split}
\label{Magnetic_energ}
\end{equation}
where $\VEC{m}_{i}$ is the direction of the magnetic moment at Gd-site $i$, 
${K}$ is a magneto-crystalline anisotropy constant and $\VEC{B}$ 
is the external magnetic field applied {along the} $c$-axis. The minimization of 
$\mathcal{E}_{m}$ is achieved by solving the Landau-Lifshitz-Gilbert 
equation as implemented in the Spirit code~\cite{Gideon:2019} (details on the construction of the phase diagram are given in~\cite{Note1}). 

The (${K},\, B$) magnetic phase diagram is presented in Fig.~\ref{panel_ph}(a). 
It displays four distinct magnetic phases: cycloidal, helical, SkX and a field-saturated ferromagnetic state.
The magnetic textures of the {first three phases} are depicted in Figs.~\ref{panel_ph}(b-d). 
{Ignoring the demagnetization field, due to the absence of DMI in centrosymmetric compounds, left- and right-handed single-$q$ helical or cycloidal states as well as lattices of N\'eel- or Bloch-type  or anti-skyrmions (ASkX), have the same energy~\cite{hoffmann2017antiskyrmions}. Experimentally, the magnetic structures can be distinguished measuring various magnetic structure factors~\cite{Note1}.}

For the case $(K, B)=(0,0)$, the cycloidal and helical phases  are 
degenerate.  For relatively low ${K}$ and moderate $B$-field, the cycloidal phase is favored. At the line $(0, B)$ we find the saturation field of about $3$T that corresponds to the unwinding energy of the cycloidal state into the field-saturated ferromagnetic one, $J(\VEC{q}_p)-J(0)$. When the value of magnetic anisotropy is increased, the cycloidal phase becomes less stable and the phase boundaries of the ferromagnetic and helical phase move towards the cycloidal phase. 
%SB keeping a low $B$-field, the helical state dominates. 
In the window $0.95\le B\le 1.4$\,T
and $0.15\le{K}\le0.45$~meV, a pocket with a 2D skyrmion lattice phase of $C_{4v}$ square symmetry emerges, which is attributed to the four-fold anisotropy in $J(\VEC{q})$ [Fig.~\ref{panel_ph}(c)].
%\SB{The critical magnetic field necessary to stabilize this skyrmion lattice phase is approximately half of the experimental value [11]. The difference is energetically a tiny quantity ($\approx 0.5$~meV) to which the choice of lattice parameters, magnetoelastic effects or higher-order interacions contribute.}
%{\color{red} The critical magnetic field necessary to stabilize this skyrmion lattice phase is approximately half of the experimental value [11], a better agreement could be achieved if multi-spin interactions would have been incorporated.}
{The critical magnetic field necessary to stabilize this skyrmion lattice phase is approximately half of the experimental value~\cite{khanh2020nanometric}. The difference is energetically small and could be due, e.g., to the precise choice of lattice parameters or small higher order interactions.}

We calculate the approximate skyrmion lattice constant to be $A_\text{Sk}\simeq~2.5$ nm, which 
is close to the {wavelength of $\lambda=2.1$\,nm we found for the $1q$} helical state in this compound. Besides the SkX phase we also found single skyrmions as meta-stable particles in the ferromagnetic phase close to the phase boundary to the cycloidal and SkX phases analogous to the discussion in Ref.\,\cite{leonov2015multiply}.

The presence of a strong magneto-elastic coupling in GdRu$_2$Si$_2$ (see Fig.~\ref{panel_1}b) 
is also reflected in the behavior of the SkX phase under pressure. To illustrate this, 
we fix $\mathcal{K}=0.3$ meV and {use} the $J_{ij}$ parameters obtained above for a $4\%$ ($6\%$) 
compression (expansion). For $4\%$ compression, the most stable state is cycloidal for 
magnetic fields up to $4$\,T, while for $6\%$ expansion the system collapses towards 
a saturated ferromagnetic state. Thus, we conclude that the SkX pocket is strongly dependent on 
the in-plane interatomic spacing. This finding is supported by experiments~\cite{Spachmann2021} 
on Gd$_2$PdSi$_3$, for which an expansion of the SkX phase was found upon uniaxial pressure.

{Our materials specific first-principles results are in accord with the theoretical model of Wang \textit{et al.}~\cite{Wang:2020} based on a 2D electron gas. Both approaches put forward a RKKY-type interaction as the principal mechanism underlying the magnetism of the centrosymmetric materials investigated. Our results are quantitatively at odds with the ab-initio results reported by Nomoto \textit{et al.}~\cite{Nomoto:2020} and hence have a different interpretation. It is apparent that the subtle long-ranged interactions of oscillatory sign and small strengths in these compounds together with the small spin-polarization of the $d$ orbitals relative to the large spin-moments associated with the Gd atoms requires a very careful assessment of all the technical or conceptional  approximations which are used in their modelling.}

\textit{Conclusion} --- Using GdRu$_2$Si$_2$ and Gd$_2$PdSi$_3$ as examples of intermetallic rare-earth compounds, we have demonstrated from first-principles that their long-range intrinsic RKKY interaction is able to generate single-$q$ magnetic spiral phases that can form skyrmion lattices with small nanometric lattice constants when laboratory magnetic fields are applied.
Single skyrmions have been found in the ferromagnetic phase near the phase boundaries to the cycloidal and skyrmion lattice.  
Our calculations directly link important skyrmion design parameters, such as the size, the symmetry, the N\'eel temperature or the applied magnetic field required to form skyrmion lattices or single skyrmions, to nesting conditions of the three-dimensional barrel-shaped Fermi surface. These can be modified and optimized in various ways,  \textit{e.g.} through the strong dependence on elastic deformations.  

The frustrated exchange introduced by the RKKY interaction opens a new vista for the stabilization of small magnetic skyrmions and skyrmion lattices in centrosymmetric compounds. Considering that GdRu$_2$Si$_2$ and Gd$_2$Pd\-Si$_3$ belong to two huge families of intermetallic compounds, there are many possibilities for material optimization.
Both the intermixing or substitution of Gd with other rare earth ions either altering the lattice parameters or increasing the crystal field effects might reduce the necessary laboratory $B$-fields. Substituting $3d$-atoms for the $4d$-ones holds the prospect of greatly raising the transition temperature and realizing room-temperature skyrmions~\cite{hou2021}.
{Indeed, from sources such as~\cite{raub1993handbook}, we can identify $225$ candidates belonging to thirteen centrosymmetric families of ternary rare-earth intermetallics with propensity for antiferromagnetic order and potential to host non-trivial spin textures~\cite{Note1}.}
Finally three-dimensional Fermi surface nesting can lead to three-dimensional modulated magnetization textures and thus a path to magnetic hopfions~\cite{rybakov2019magnetic}.

\begin{acknowledgements}
We thank Nikolai Kiselev, Phivos Mavropoulos, Markus Hoffmann and Moritz Sallermann for fruitful discussions. 
S.B.\ gratefully acknowledges financial support from the 
European Research Council (ERC) under the European Union's Horizon 
2020 research and innovation program (Grant No.\ 856538, project ``3D MAGiC''), from Deutsche For\-schungs\-gemeinschaft 
(DFG) through SPP 2137 ``Skyrmionics'' (Project BL 444/16) and the Collaborative 
Research Centers SFB 1238 (Project C01), respectively. J.B.S.\ acknowledges support from UK EPSRC Grant No.\ EP/M028941/1. 
J.B. and S.B.\ gratefully acknowledge the computing time granted by the JARA-CSD and VSR Resource Allocation Board provided on the supercomputers CLAIX at RWTH Aachen University and JURECA at Supercomuter Centre J\"ulich under grant nos.\ jara0219, jara3dmagic. 
\end{acknowledgements}

%%%%%%%%%%%%%%%%%%%%%%%%%%%%%%%%
%%%%%%%%% Bibliography %%%%%%%%%
%%%%%%%%%%%%%%%%%%%%%%%%%%%%%%%%
\bibliography{bibliography.bib}
%%%%%%%%%%%%%%%%%%%%%%%%%%%%%%%%

\end{document}